\documentstyle [12pt] {article}
\begin{document}

%%%%Start of Text%%%%%%%%%%%%%%%%%%%%%%%%%%%%%%%%%%%%%%%%%%%%%%%%%%%%%%%%%%%%
\pagestyle{empty}
\rightline{\vbox{
\halign{&#\hfil\cr
&NUHEP-TH-95-1 \cr
&January 1995 \cr
&hep-ph/9501296 \cr }}}
\bigskip
\bigskip
\bigskip
{\Large\bf
	\centerline{QCD Radiative Corrections to the}
	\centerline{Leptonic Decay Rate of the $B_c$ Meson}}
\bigskip
\normalsize

\centerline{Eric Braaten and Sean Fleming}
\centerline{\sl Department of Physics and Astronomy, Northwestern University,
    Evanston, IL 60208}
\bigskip

\begin{abstract}
The QCD radiative corrections to the leptonic decay rate of the $B_c$ meson
are calculated using the formalism of nonrelativistic QCD (NRQCD) to separate
short-distance and long-distance effects.
The $B_c$ decay constant is factored into a sum of
NRQCD matrix elements each multiplied by a short-distance coefficient.
The short-distance coefficient for the leading matrix element
is calculated to order $\alpha_s$
by matching a perturbative calculation in full QCD with the corresponding
perturbative calculation in NRQCD.  This short-distance correction
decreases the leptonic decay rate by approximately $15\%$.
\end{abstract}

\vfill\eject\pagestyle{plain}\setcounter{page}{1}

The study of heavy quarkonium systems has
played an important role in the development of quantum
chromodynamics (QCD). Some of the earliest applications of perturbative QCD
were
calculations of the decay rates of charmonium \cite{ap}.  These
calculations were based on the assumption that, in the nonrelativistic limit,
the decay rate factors into a short-distance
perturbative part associated with the annihilation of the heavy quark and
antiquark and a long-distance part associated with the
quarkonium wavefunction.  This simple factorization
assumption fails for P-wave states \cite{barb}, which satisfy a more
general factorization formula containing
two nonperturbative factors \cite{bbl1}.
Calculations of the annihilation decay rates of heavy quarkonium
have recently been placed on a solid theoretical foundation by
Bodwin, Braaten, and Lepage \cite{bbl}.
Their approach is based on nonrelativistic QCD
(NRQCD), an effective field theory that is equivalent to QCD to any given
order in the relative velocity $v$ of the heavy quark and antiquark \cite{cl}.
Using NRQCD to separate the short-distance and
long-distance effects, Bodwin, Braaten, and Lepage derived
a general factorization formula for the inclusive annihilation
decay rates of heavy quarkonium.
The short-distance factors in the factorization formula
can be calculated using perturbative QCD, and the
long-distance factors are defined rigorously in terms of matrix elements
of NRQCD that can be evaluated using lattice calculations.
The general factorization formula applies equally well to S-waves, P-waves,
and higher orbital-angular-momentum states, and it can be used
to systematically incorporate relativistic corrections to the decay rates.

Since the top quark decays too quickly to produce narrow resonances,
the only heavy quark-antiquark bound states that remain to be discovered
are the ${\bar b} c$ mesons and their antiparticles.
The possibility that ${\bar b} c$ mesons may be discovered at
existing accelerators has stimulated much recent work on the
properties of these mesons \cite{eq} and on their production cross
sections at high energy colliders \cite{prod}.
Once produced, a ${\bar b} c$ meson will cascade down through
lower energy ${\bar b} c$ states via hadronic or electromagnetic
transitions to the pseudoscalar ground state $B_c$ which decays weakly.
The discovery of the $B_c$ meson will require a detailed understanding
of its decay modes.
In this paper, we compute the short-distance QCD radiative correction
to the leptonic decay rate of the $B_c$.
We use the formalism of NRQCD to factor the amplitude for the decay
into short-distance coefficients multiplied by NRQCD
matrix elements. The short-distance coefficient for the leading
matrix element is calculated to next-to-leading order in $\alpha_s$
by matching a perturbative calculation in full QCD with the corresponding
perturbative calculation in NRQCD.

The leptonic decay of the $B_c$ proceeds through a virtual $W^+$
as in Figure 1.  The $W^+$ couples to the $B_c$ through the axial-vector
part of the charged weak current.  All QCD effects, both perturbative
and nonperturbative, enter into the decay rate through the decay constant
$f_{B_c}$, defined by the matrix element
\begin{equation}
\langle 0 | {\bar b} \gamma^\mu \gamma_5 c | B_c(P) \rangle
\; = \; i f_{B_c} P^\mu ,
\label{fbc}
\end{equation}
where $| B_c(P) \rangle$ is the state consisting of a $B_c$
with four-momentum $P$.  It has the standard covariant normalization
$\langle B_c(P') | B_c(P) \rangle = (2 \pi)^3 2 P^0 \delta^3(P' - P)$
and its phase has been chosen so that $f_{B_c}$ is real and positive.
In terms of the decay constant $f_{B_c}$, the leptonic decay rate is
\begin{equation}
\Gamma(B_c \to \ell^+ \nu_\ell) \;=\;
{1 \over 8 \pi} |V_{bc}|^2 G_F^2 M_{B_c} f_{B_c}^2 m_\ell^2
	\left( 1 - {m_\ell^2 \over M_{B_c}^2} \right)^2 ,
\end{equation}
where $V_{bc}$ is the appropriate Kobayashi-Maskawa matrix element,
$G_F$ is the Fermi constant, $M_{B_c}$ is the mass of the $B_c$ meson,
and $m_\ell$ is the mass of the charged lepton.

The formula (\ref{fbc}) provides a nonperturbative definition of the
decay constant $f_{B_c}$, so that it can be calculated using lattice
QCD simulations.  One of the difficulties with such a calculation
is that it requires a lattice with large volume and fine lattice spacing,
since the strong interactions must be accurately simulated over
many distance scales.  The long-distance scales range from
$1/\Lambda_{QCD}$, the scale of nonperturbative effects associated
with gluons and light quarks, to the scale $1/(m_c v)$
of the meson structure, where $v$ is the typical relative velocity
of the charm quark.
The short-distance scales include the Compton wavelengths $1/m_c$ and $1/m_b$
of the heavy quark and antiquark.  A more effective strategy for calculating
$f_{B_c}$ is to separate short-distance effects from long-distance effects,
to calculate the short distance effects analytically using perturbation theory
in $\alpha_s$, and to use lattice simulations only for calculating
the long-distance effects.  Having already taken into account
the short-distance effects,
one can use a much coarser lattice which provides enormous savings in computer
resources.

An elegant way to separate short-distance and long-distance effects is to
use NRQCD, an effective field theory in which heavy quarks
are described by a nonrelativistic Schroedinger field theory
of 2-component Pauli spinors.  The lagrangian is
\begin{equation}
{\cal L}_{\rm NRQCD}
\;=\; {\cal L}_{\rm light}
\;+\; \psi_c^\dagger \left(i D_0 + {\bf D}^2/(2 m_c) \right) \psi_c
\;+\; \chi_b^\dagger \left(i D_0 - {\bf D}^2/(2 m_b) \right) \chi_b
\;+\; \ldots ,
\label{NRQCD}
\end{equation}
where ${\cal L}_{\rm light}$ is the usual relativistic lagrangian for
gluons and light quarks.  The 2-component field $\psi_c$ annihilates
charm quarks, while $\chi_b$ creates bottom antiquarks.
The typical velocity $v$ of the charm quark in the meson provides
a small parameter that can be used as a nonperturbative expansion parameter.
Lattice simulations using the terms given
explicitly in (\ref{NRQCD}) can be used to calculate matrix elements
for the $B_c$ with errors of order $v^2$.  To obtain higher accuracy,
additional terms, represented by $\ldots$ in (\ref{NRQCD}),
must be included.  There are 8 terms that must be added to decrease
the errors to order $v^4$.

To express the decay constant $f_{B_c}$ in terms of NRQCD matrix elements,
we must express the axial-vector current ${\bar b} \gamma^\mu \gamma_5 c$
in terms of NRQCD fields.  Only the $\mu = 0$ component contributes to the
matrix element (\ref{fbc}) in the rest frame of the $B_c$.  This
component of the current has an operator expansion in terms of NRQCD fields:
\begin{equation}
{\bar b} \gamma^0 \gamma_5 c
\;=\; C_0(m_b,m_c) \; \chi_b^\dagger \psi_c
\;+\; C_2(m_b,m_c) \; ({\bf D} \chi_b)^\dagger \cdot {\bf D} \psi_c
\;+\; \ldots,
\label{opex}
\end{equation}
where $C_0$ and $C_2$ are short-distance coefficients that depend on
the quark masses $m_b$ and $m_c$.  By dimensional analysis, the coefficient
$C_2$ is proportional to $1/m_Q^2$.  The contribution to the matrix element
$\langle 0 | {\bar b} \gamma^0 \gamma_5 c | B_c \rangle$ from the operator
$({\bf D} \chi_b)^\dagger \cdot {\bf D} \psi_c$
is suppressed by $v^2$ relative to the operator
$\chi_b^\dagger \psi_c$, where $v$ is the
typical velocity of the charm quark in the $B_c$.
The $\ldots$ in (\ref{opex}) represent other operators whose contributions
are suppressed by higher powers of $v^2$.

The short-distance coefficient $C_0$ and $C_2$ can be determined
by matching perturbative calculations of the matrix elements in
full QCD and NRQCD.  A convenient choice for matching is the matrix element
between the vacuum and the state $|c {\bar b} \rangle$
consisting of a c and a ${\bar b}$ on their perturbative mass shells
with nonrelativistic four-momenta $p$ and $p'$ in the center of momentum
frame:  ${\bf p} + {\bf p}' = 0$.  The matching condition is
\begin{equation}
\langle 0 | {\bar b} \gamma^0 \gamma_5 c
	| c {\bar b} \rangle \Bigg|_{\mbox{{\scriptsize P-QCD}}}
= C_0 \; \langle 0 | \chi_b^\dagger \psi_c
	| c {\bar b} \rangle \Bigg|_{\mbox{{\scriptsize P-NRQCD}}}
 \;+\; C_2 \; \langle 0 | ({\bf D} \chi_b)^\dagger \cdot
 {\bf D} \psi_c | c {\bar b}\rangle \Bigg|_{\mbox{{\scriptsize P-NRQCD}}}
\;+\; \ldots,
\label{match}
\end{equation}
where P-QCD and P-NRQCD represent perturbative QCD and perturbative NRQCD,
respectively.
At leading order in $\alpha_s$, the matrix element on the left side of
(\ref{match}) is ${\bar v}_b(-{\bf p}) \gamma^0 \gamma_5 u_c({\bf p})$.
The Dirac spinors are
\label{spinor}
\begin{eqnarray}
u_c(\mbox{{\bf p}}) &=& \sqrt{E_c+m_c \over 2E_c} \left(
\begin{array}{c}
	\xi \\
	{\mbox{{\bf p}} \cdot \mbox{{\boldmath $\sigma$}}
		\over E_c+m_c} \xi
\end{array}
	\right) ,
\label{uspinor}
\\
v(-\mbox{{\bf p}}) &=& \sqrt{E_b+m_b \over 2E_b}	\left(
\begin{array}{c}
	{(-\mbox{{\bf p}}) \cdot \mbox{{\boldmath $\sigma$}}
		\over E_b+m_b} \eta \\
	\eta
\end{array}
	\right) ,
\label{vspinor}
\end{eqnarray}
where $\xi$ and $\eta$ are 2-component spinors
and $E_Q = m_Q^2 + {\bf p}^2$.  Making a nonrelativistic expansion of
the spinors to second order in ${\bf p}/m_Q$, we find
\begin{equation}
{\bar v}_b(-{\bf p}) \gamma^0 \gamma_5 u_c({\bf p})
\;\approx\; \eta_b^\dagger \xi_c
\left( 1 \;-\; {1 \over 8} \left( {m_b + m_c \over m_b m_c} \right)^2 {\bf p}^2
	\;+\; \ldots \right).
\end{equation}
At leading order in $\alpha_s$,
the matrix elements on the right side of (\ref{match}) are
$\eta_b^\dagger \xi_c$ and ${\bf p}^2 \eta_b^\dagger \xi_c$.
The short distance coefficients are therefore
$C_0 = 1$ and
\begin{equation}
C_2 \;=\; - {1 \over 8 \; m_{\rm red}^2} ,
\label{C2}
\end{equation}
where $m_{\rm red} = m_b m_c/(m_b+m_c)$ is the reduced mass.

To determine the short distance coefficients to order $\alpha_s$,
we must calculate the matrix elements on both sides of (\ref{match})
to order $\alpha_s$.  We will calculate the order-$\alpha_s$ correction
only for the coefficient $C_0$, since the contribution proportional
to $C_2$ is suppressed by $v^2$.
The coefficient $C_0$ can be isolated by taking the
limit ${\bf p} \to 0$, in which case the matrix element
of $({\bf D} \chi_b)^\dagger \cdot {\bf D} \psi_c$ vanishes.
We first calculate the matrix element on the left side of (\ref{match})
to order $\alpha_s$.  The relevant diagrams are shown in Fig.~2.
The tree diagram in Fig.~2a gives a product of Dirac spinors
${\bar v}_b \gamma_0 \gamma_5 u_c$.
We calculate the loop diagrams in Fig.~2b-2d in Feynman gauge,
using dimensional regularization
to regularize both infrared and ultraviolet divergences.
Momentum integrals are analytically continued to
$D = 4 - 2 \epsilon$ spacetime dimensions, requiring the introduction
of a regularization scale $\mu$.
The effects of the quark self-energy diagrams in Fig.~2b and 2c
is to multiply the matrix element by the renormalization constants
$\sqrt{Z_b}$ and $\sqrt{Z_c}$, where
\begin{equation}
\sqrt{Z_Q}
\;=\; 1 \;+\; {2 \alpha_s \over 3 \pi}
	\left( - {1 \over 4} {1 \over \epsilon_{UV}}
		- {1 \over 2} {1 \over \epsilon_{IR}}
		- {3 \over 4} (\log 4 \pi - \gamma)
		+ {3 \over 2} \log{m_Q \over \mu} - 1 \right) .
\end{equation}
The subcript $UV$ or $IR$ on the $\epsilon$'s indicates whether the
divergence is of ultraviolet or infrared origin.
Combining the propagators using the Feynman parameter trick and
using the fact that the external quarks ${\bar b}$ and $c$ are on-shell,
the vertex correction from Fig.~2d can be reduced to the tree-level
diagram in Fig.~2a multiplied by the factor
\begin{eqnarray}
\Lambda  \; = \;  {64 \pi i \alpha_s \over 3} \!\! & \mbox{} & \!\!\!\!\!\!
\int_0^1 dx \int_0^{1-x} dy  \; \mu^{2\epsilon} \int {d^Dk \over (2 \pi)^D}
	{1 \over [k^2 - (x p' - y p)^2 + i \epsilon ]^3 }
\nonumber \\
 && \Bigg[ 2 (1-x-y) p \cdot p' + x m_c^2 + y m_b^2 + (x+y) m_b m_c
\nonumber \\
&& \;\; + 2 (1 - \epsilon) (x m_c - y m_b)^2 - (1-\epsilon) (x p' - y p)^2
- {(1-\epsilon)^2 \over (2-\epsilon)} k^2 \Bigg] .
\end{eqnarray}
After integrating over $k$,
it is convenient to change variables from the Feynman parameters
to $s = x+y$ and $t = x/s$.  The $s$-integrals are trivial, but the
$t$-integrals must be evaluated with care using the $i \epsilon$ prescription
in the denominator.  Setting $p \cdot p' \approx m_b m_c (1 + v^2/2)$,
where $v$ is the relative velocity of the ${\bar b}$ and $c$, and taking
the limit $v \to 0$, the integral reduces to
\begin{eqnarray}
\Lambda \;=\;  {2 \alpha_s \over 3 \pi}
\left[ {1 \over 2} {1 \over \epsilon_{UV}} + {1 \over \epsilon_{IR}}
	+ {3 \over 2} (\log 4 \pi - \gamma) \right.
        \!\! &+& \! \!
        {\pi^2 \over v} - 3 {m_b \over m_b + m_c} \log {m_c \over \mu}
	- 3 {m_c \over m_b + m_c} \log {m_b \over \mu} - 1
        \nonumber \\ && \left.
	- {i \pi \over v} \left( {1 \over \epsilon_{IR}}
		- 2 \log{2 m_{\rm red} v \over \mu}
		+ \log 4 \pi - \gamma \right) \right],
\end{eqnarray}
where $m_{\rm red} = m_b m_c/(m_b+m_c)$.
Multiplying the tree-level matrix element by the
vertex factor $1 + \Lambda$ and by the renormalization constants
$\sqrt{Z_b}$ and $\sqrt{Z_c}$, we obtain the final answer for the matrix
element to order $\alpha_s$:
\begin{eqnarray}
\langle 0 | {\bar b} \gamma^0 \gamma_5 c | c {\bar b}  \rangle
\Bigg|_{\mbox{{\scriptsize P-QCD}}}
\;=\; {\bar v}_b \gamma^0 \gamma^5 u_c
\Bigg\{ 1 &+& {2 \alpha_s \over 3 \pi}
	\Bigg[ {\pi^2 \over v}
		+ {3 \over 2} {m_b-m_c \over m_b+m_c} \log {m_b \over m_c}
		- 3
\nonumber \\
	       &&- {i \pi \over v} \left( {1 \over \epsilon_{IR}}
			- 2 \log {2 m_{\rm red} v \over \mu} + \log{4 \pi}
                        -\gamma \right)
		 \Bigg] \Bigg\} .
\label{meQCD}
\end{eqnarray}
Note that the ultraviolet divergences have cancelled.
The imaginary part of (\ref{meQCD}) arises because it is
possible for the quark and antiquark created by the current to scatter
on-shell.

We next compute the NRQCD matrix element on the right side of (\ref{match})
to order $\alpha_s$. The relevant Feynman diagrams are again shown in Fig.~2.
The tree diagram in Fig.~2a gives a product of Pauli spinors
$\eta_b^\dagger \xi_c$.  It is convenient to calculate the loop diagrams in
Coulomb gauge.  In this gauge,
the coupling of transverse gluons to heavy quarks is proportional
to the heavy quark velocity, and therefore does not contribute
in the limit $v \to 0$.  We therefore need only calculate the contribution
from Coulomb exchange.  The wavefunction renormalization factors associated
with the diagrams in Fig.~2b and Fig.~2c are trivial:
$\sqrt{Z_b} = \sqrt{Z_c} = 1$.
The diagram in Fig.~2d reduces to a multiplicative
correction to the tree-level matrix element from Fig.~2a:
\begin{equation}
\Lambda \;=\; - {16 \pi i \alpha_s \over 3}
\; \mu^{2 \epsilon} \int{d^Dq \over (2 \pi)^D} \; {1 \over {\bf q}^2}
	{1 \over E_b + q_0 - ({\bf p} + {\bf q})^2/2m_b + i \epsilon} \;
	{1 \over E_c - q_0 - ({\bf p} + {\bf q})^2/2m_c + i \epsilon} \;,
\label{Lam}
\end{equation}
where $E_Q = p^2/2m_Q$.
The infrared-divergent integral has been dimensionally regularized.
After using contour integration to integrate over the energy $q_0$ of
the exchanged gluon, we find that (\ref{Lam}) reduces to an
integral over the gluon's three-momentum:
\begin{equation}
\Lambda \;=\; {32 \pi \alpha_s m_{\rm red} \over 3} \; \mu^{2\epsilon}
\int{d^{3-2\epsilon}q \over (2 \pi)^{3-2\epsilon}} {1 \over {\bf q}^2}
	{1 \over {\bf q}^2 + 2 {\bf p} \cdot {\bf q} - i \epsilon} \;.
\label{vNRQCD}
\end{equation}
Evaluating the regularized integral in (\ref{vNRQCD}), we obtain
\begin{equation}
\langle 0 | \chi_b^\dagger \psi_c | c {\bar b} \rangle
\Bigg|_{\mbox{{\scriptsize P-NRQCD}}}
\;=\; \eta_b^\dagger \xi_c
\Bigg\{ 1 + {2 \alpha_s \over 3 \pi}
\Bigg[ {\pi^2 \over v}
\;-\; {i \pi \over v} \left( {1 \over \epsilon_{IR}}
	- 2 \log{2 m_{\rm red} v \over \mu}
	+ \log 4 \pi  - \gamma \right) \Bigg] \Bigg\} \;,
\label{meNRQCD}
\end{equation}
where $v$ is the relative velocity of the $\bar b$ and $c$:
$|{\bf p}| = m_{\rm red} v$.
Note that the infrared divergences and the factors of $1/v$ are
identical to those in the matrix element (\ref{meQCD}) for full QCD.

Comparing the matrix elements (\ref{meQCD}) and (\ref{meNRQCD}),
we can read off the short-distance coefficient $C_0$
from the matching condition (\ref{match}):
\begin{equation}
C_0 \;=\;
1 \;+\; {\alpha_s(m_{\rm red}) \over \pi}
	\left[{m_b-m_c \over m_b+m_c} \log {m_b \over m_c}
		- 2 \right] .
\label{C0}
\end{equation}
We have chosen $m_{\rm red}$ as the scale of the running coupling constant.
To the accuracy of this calculation,
any scale of order $m_b$ or $m_c$ would be equally correct.
Setting $m_b = 4.5$ GeV and $m_c = 1.5$ GeV, we find that
$\alpha_s(m_{\rm red}) \approx 0.34$,
and $C_0 \approx 0.85$.

Our final result for the decay constant of the $B_c$ is
\begin{equation}
i f_{B_c} M_{B_c} \;=\; C_0 \; \langle 0 | \chi_b^\dagger \psi_c | B_c \rangle
\;+\; C_2 \; \langle 0 | ({\bf D} \chi_b)^\dagger \cdot
	{\bf D} \psi_c | B_c \rangle .
\end{equation}
The short-distance coefficient $C_0$ is given to next-to-leading order
in $\alpha_s$ in (\ref{C0}),
while $C_2$ is given to leading order in (\ref{C2}).
The uncertainties consist of the perturbative errors in the short distance
coefficients and an error of relative order $v^4$ from
the neglect of matrix elements that are higher order in $v^2$.
To achieve an error of order $v^4$, the matrix element
$\langle 0 | ({\bf D} \chi_b)^\dagger \cdot {\bf D} \psi_c | B_c \rangle$
can be calculated by lattice simulations using those terms in the NRQCD
lagrangian that are given explicitly in (\ref{NRQCD}), but
$\langle 0 | \chi_b^\dagger \psi_c | B_c \rangle$
must be calculated using an improved action.
If an error of order $v^2$ is sufficient
accuracy, than the matrix element
$\langle 0 | ({\bf D} \chi_b)^\dagger \cdot {\bf D} \psi_c | B_c \rangle$
can be dropped, and $\langle 0 | \chi_b^\dagger \psi_c | B_c \rangle$
can be calculated using only those terms in the NRQCD
lagrangian that are given explicitly in (\ref{NRQCD}).
The parameters in this lagrangian
are $\alpha_s$ and the quark masses $m_b$ and $m_c$, all of which can
be tuned so as to reproduce the spectroscopy of charmonium and bottomonium.
These simulations should be very accurate for the intermediate case of
the ${\bar b} c$ system.
In the absence of any lattice calculations of the ${\bar b} c$ system,
the matrix element $\langle 0 | \chi_b^\dagger \psi_c | B_c \rangle$
can be estimated using wavefunctions at the origin from
nonrelativistic potential models:
\begin{equation}
|\langle 0 | \chi_b^\dagger \psi_c | B_c \rangle|^2
\;\approx\; 2 M_{B_c} {3 \over 2 \pi} |R(0)|^2 .
\end{equation}
The factor of $2 M_{B_c}$ takes into account the relativistic normalization of
the state $|B_c \rangle$.

Although it may not be relevant to the $B_c$ meson, it is interesting
to consider the limit $m_b \gg m_c$.  In this limit,
the perturbative correction in (\ref{C0}) contains a large
logarithm of $m_b/m_c$.  Heavy Quark Effective Theory can be used
to sum up the leading logarithms of $m_b$ \cite{HQET}:
\begin{equation}
C(\alpha_s) \;\longrightarrow\;
\left( {\alpha_s(\Lambda) \over \alpha_s(m_b)} \right)^{6\pi/(33-2n_f)}
\left\{ 1 \;+\; {\alpha_s(m_c) \over \pi}
	\left[ \log {\Lambda \over m_c} - 2 \right] \right\} ,
\end{equation}
where $n_f$ is the number of light flavors, including the charm quark,
and where $\Lambda$ is an arbitrary factorization scale.

We have calculated the short-distance QCD radiative correction to the
leptonic decay rate of the $B_c$ meson.
This calculation provides an illustration of the factorization methods
based on NRQCD that were developed in Ref. \cite{bbl}.
The decay constant $f_{B_c}$ of the $B_c$ was factored
into short-distance coefficients multiplied by NRQCD matrix elements.
The radiative correction to the coefficient of the leading matrix element
decreases the leptonic decay rate of the $B_c$ meson by about $15 \%$.

This work was supported in part by the U.S. Department of Energy,
Division of High Energy Physics, under Grant DE-FG02-91-ER40684.

\bigskip
\noindent{\Large\bf Figure Captions}
\begin{enumerate}
\item Diagram for the annihilation of $B_c$ into lepton pairs via a
virtual $W^+$.  The shaded oval represents the wavefunction of the $B_c$.
\item Diagrams for the matrix elements
$\langle 0 | {\bar b} \gamma^0 \gamma_5 c | c {\bar b} \rangle$
in perturbative QCD and
$\langle 0 | \chi_b^\dagger \psi_c | c {\bar b} \rangle$
in perturbative NRQCD.
\end{enumerate}
\vfill\eject

\end{document}